\documentclass[bibyear]{aa}
\usepackage{graphicx}
\usepackage{amssymb}
\usepackage{tabulary}
\usepackage{natbib,twoopt}

\begin{document}

\title{Prospects for studying million-degree gas in the Milky Way halo using 
the forbidden optical [Fe\,{\sc x}] and [Fe\,{\sc xiv}] intersystem lines}

   \author{
          P. Richter \inst{1},
          F. R\"unger  \inst{1},
          N. Lehner \inst{2},
          J.C. Howk \inst{2},
          C. P\'eroux \inst{3,4},
          N. Libeskind \inst{5},
          M. Steinmetz \inst{5},
       \and
          R. de Jong \inst{5}
          }
   \offprints{P. Richter\\
   \email{prichter@astro.physik.uni-potsdam.de}}

   \institute{Institut f\"ur Physik und Astronomie, Universit\"at Potsdam,
             Karl-Liebknecht-Str.\,24/25, 14476 Golm, Germany
   \and
   Department of Physics and Astronomy, University of Notre Dame, Notre Dame, IN 46556, USA
   \and
   European Southern Observatory (ESO), Karl-Schwarzschildstrasse 2, D-85748 Garching, Germany
   \and
   Aix Marseille Universit\'e, CNRS, LAM (Laboratoire d’Astrophysique de Marseille) UMR 7326, F-13388 Marseille, France
   \and
   Leibniz-Institut für Astrophysik Potsdam (AIP), An der Sternwarte 16, 14482 Potsdam, Germany
   }

\date{Received 2 June 2025; accepted 16 July 2025}


\abstract
{
The Milky Way is surrounded by large amounts of hot gas at
temperatures $T>10^6$ K, which represents a major baryon reservoir.
}
{
We explore the prospects of studying the hot coronal gas in 
Milky Way halo by analyzing the highly forbidden optical coronal
lines of [Fe\,{\sc x}] and [Fe\,{\sc xiv}] in absorption against
bright (unrelated) extragalactic background sources. 
}
{
We use a semi-analytic model of the Milky Way's coronal gas distribution 
together wih HESTIA simulations of the Local Group and observational constraints to predict the 
expected Fe\,{\sc x} and Fe\,{\sc xiv} column densities as well as the line shapes 
and strengths for the [Fe\,{\sc x}] $\lambda 6374.5$ and
[Fe\,{\sc xiv}] $\lambda 5302.9$ transitions. We provide predictions for
the signal-to-noise (S/N) ratio required to detect these lines. Using archical optical data from
an original sample of 739 high-resolution AGN spectra from VLT/UVES and 
KECK/HIRES, we generate a stacked composite spectrum to measure
an upper limit for the column densities of Fe\,{\sc x} and Fe\,{\sc xiv} in 
the Milky Way's coronal gas.
}
{
We predict column densities of log $N$(Fe\,{\sc x}$)= 15.40$ and
log $N$(Fe\,{\sc xiv}$)=15.23$ in the Milky Way's hot halo, corresponding to
equivalent widths of $W_{\rm FeX,6347}=190\,\mu$\AA\,and $W_{\rm FeXIV,5302}=220\,\mu$\AA.
We estimate that a minimum S/N of $\sim 50,000$ ($\sim 25,000$) is required to detect 
[Fe\,{\sc x}] $\lambda 6374.5$ ([Fe\,{\sc xiv}] $\lambda 5302.9$) absorption at a $3\sigma$ level.
No [Fe\,{\sc x}] and [Fe\,{\sc xiv}] is detected in our composite spectrum,
which achieves a maximum S/N$=1,240$ near 5300 \AA. We derive $3\sigma$ upper
column-density limits of log $N$(Fe\,{\sc x}$)\leq 16.27$ and log $N$(Fe\,{\sc xiv}$)\leq 15.85$, 
in line with the above-mentioned predictions.
}
{
While [Fe\,{\sc x}] and [Fe\,{\sc xiv}] absorption is too weak to be detected with
current optical data, we outline how up-coming extragalactic spectral surveys with millions
of medium- to high-resolution optical spectra will provide the necessary sensitivity
and spectral resolution to measure velocity-resolved [Fe\,{\sc x}] and [Fe\,{\sc xiv}] 
absorption in the Milky Way's
coronal gas (and beyond). This gives the prospect of opening a new window for 
studying the dominant baryonic mass component of the Milky Way in the form
of hot coronal gas via optical spectroscopy.
}

\authorrunning{Richter et al.}
\titlerunning{Optical absorption in million-degree gas in the Milky Way halo}

\maketitle

%


\section{Introduction}

The dominant baryonic mass reservoir in galaxies is the
diffuse gas contained within their halos. This so-called 
circumgalactic medium (CGM) plays a crucial role for the on-going
formation and evolution of galaxies and thus has become a major
research field in modern astrophysics \citep{tumlinson2017}.

As a consequence of the highly complex physics (including the full
magneto-hydrodynamics, cosmic rays, non-equilibrium cooling and
heating) and the dynamics of 
the processes that contribute to the CGM's mass and energy
budget, the CGM is an extremely heterogeneous gaseous medium with
multiple phases. In a simplified picture, the CGM is composed of 
cool and warm ($T=10^2-10^5$ K) streams of gas that are embedded 
in a hotter gaseous medium ($T=10^6-10^7$ K).
As observations in the Milky Way's CGM show, the sizes of circumgalactic gas 
features range from AU scales \citep[in high-density clumps;][]{richter2003,meyer1999} 
to several hundred kpc 
\citep[in large, coherent gas streams such as the Magellanic Stream;][]{putman2003,richter2017b}.
Only in the Milky Way can the CGM be observationally resolved at all
these scales, making our Galaxy an important test ground for our
general understanding of the circumgalactic medium. It is, on the other hand, 
this enormous dynamic range in physical conditions and spatial scales that makes 
it so challenging for magneto-hydrodynamical simulations to provide a 
realistic picture of CGM in all its facets \citep[see][and references therein]{crain2023}.

The global CGM properties of Milky Way type galaxies can be modeled in the 
context of $\Lambda$CDM galaxy formation models \citep{white1991,naab2017}, where it is 
predicted that diffuse gas in cosmological filaments is accreted onto dark matter 
halos and shock-heated to approximately the halo virial temperature (a few 
million degrees for Milky-Way type galaxies). This hot gas is sometimes referred to 
as ``galactic corona'', in analogy to the Sun's hot coronal gas \citep{spitzer1956}.
Cooling processes from the coronal gas, the stripping of satellite galaxies,
and the accretion of metal-poor gas from the intergalactic medium (IGM) 
lead to the formation cold gas streams that move towards the disk where the gas
is cooled and transformed into stars or is fueling the coronal gas of more massive galaxies
\citep{afruni2023}. In addition to accretion of material from the corona and
from outside, gas (and energy) is also ejected into the CGM from inside from
the disk as a result of feedback processes \citep{marasco2022} after which it
is re-accreted onto the disk. Therefore, the CGM is believed to represent a 
substantial baryon reservoir that reflects the evolutionary state of its host 
galaxy \citep{peroux2020}.

With its low gas densities (log $n_{\rm H}=-2$ to $-5$, typically) and 
high temperatures ($T>10^6$ K), the hot coronal gas phase is particularly
difficult to observe \citep{comparat2022}. For the Milky Way, the X-ray transitions 
of O\,{\sc vii}, O\,{\sc viii} and other high ions provide important diagnostic 
lines to study the Galaxy's hot coronal gas component. These lines
can be observed either in absorption against X-ray bright active galactic nuclei (AGN)
or in emission \citep{fang2006,miller2013,miller2015,hodges2016,li2017,das2021}.
Recent X-ray studies indicate that the Milky Way's hot coronal gas 
contains a total mass as much as $\sim 10^{11} M_{\sun}$ 
within $250$ kpc \citep[e.g.,][]{miller2013,miller2015}.
A major drawback of CGM X-ray observations is, however, the very low
spectral resolution of the data that precludes a precise localization
of the hot gas resevoirs within the Milky Way, its circumgalactic
environment and in the Local Group \citep{bhattacharyya2023}, giving 
only limited information on its overall dynamics.

Unlike in the X-ray band, there are no strong resonance
lines from highly-ionized metals available in the ultraviolet (UV)/optical regime that 
would independently trace million-degree gas in the CGM and IGM at $z=0$
in a direct manner. 
Indirect methods to trace hot gas in the local Universe in the CGM and IGM via 
intermediate-and high-resolution UV absorption 
spectroscopy include the measurements of transition-temperature gas 
($T\sim 10^{5.5}$ K) using the O\,{\sc vi} doublet 
\citep{sembach2003,savage2003,tripp2008}
and the analysis of thermally broadened
H\,{\sc i} Ly\,$\alpha$ absorption lines (BLAs) and Coronal BLAs (CBLAs)
that trace the tiny neutral gas fraction in the hot
medium \citep{richter2006a,richter2006b,lehner2007,tepper2012,richter2020}.
For the study of the CGM of the Milky Way and other Local Group member galaxies, the 
analysis of CBLAs is not an option, however, as the damped interstellar
H\,{\sc i} Ly\,$\alpha$ absorption superposes any CBLA signature within
$|v_{\rm LSR}|\approx 700$ km\,s$^{-1}$. High-ion transitions in the
extreme UV, such as Ne\,{\sc viii} $\lambda\lambda 770.4,780.3$, cannot be used
to trace hot gas in the $z=0$ Universe, but are accessible only at higher 
redshifts \citep{tepper2014}.

A potential alternative to strong resonance lines are (exceedingly weak) forbidden 
transitions of high-ions, the most prominent of them (from early studies of the Sun) being 
located in the optical regime. Already in the 1980s, \citet{york1983}
discussed the possibility to use the intersystem lines of
[Fe\,{\sc x}] near $6375$ \AA\, and [Fe\,{\sc xiv}] near $5303$ \AA\, as
the most promising candidate lines from the set of forbidden transitions
in the optical regime to sample shock-heated, hot gas in the CGM and IGM in optical 
spectra from ground-based telescopes. However, 
because of the very small oscillator strengths of these
forbidden transitions, an extremely high S/N of a few thousand would
be required to detect these lines in the spectra of background AGN,
which is (so far) not feasible for individual targets.
In collisional ionization equilibrium, 
these lines trace gas at temperatues $T=10^{6.0}$-$10^{6.3}$ K (see Fig.\,1), 
thus close to the halo virial temperature of
Milky-Way type galaxies (see above). In the 1980s, several groups 
searched for [Fe\,{\sc x}] at $\lambda 6374.5$ and
[Fe\,{\sc xiv}] $\lambda 5302.9$ absorption in the Milky Way ISM and 
in hot gas towards the LMC 
\citep{hobbs1984a,hobbs1984b,hobbs1985,pettini1986,wang1989,pettini1989,malaney1988},
but none found a convincing signal (despite the inital claims of a detection; 
see discussion in Sect.\,2.2). 
In complementing these optical studies, \citet{anderson2016,anderson2018} have 
investigated emission signatures of hot gas around the nucleus of M87 using
the highly forbidden transition of [Fe\,{\sc xxi}] in the far UV at $1354.1$ \AA.

In \citet{richter2014}, we discussed the possibility of using the 
[Fe\,{\sc x}] $\lambda 6374.5$ and [Fe\,{\sc xiv}] $\lambda 5302.9$ lines
in stacks of archival optical spectra of extragalactic sightlines from VLT/UVES 
to put constraints on the hot gas mass in the Milky Way coronal gas. 
There are several compelling reasons to seriously consider these 
transitions for a systematic study of the Milky Way's hot coronal gas:
i) detecting [Fe\,{\sc x}] and/or [Fe\,{\sc xiv}] would offer independent 
constraints on the mass of hot gas in and around the Milky Way, complementing 
other measurement methods; ii) acquiring the necessary data incurs no additional 
cost, as they will be readily available from numerous ongoing and upcoming 
extragalactic spectral surveys; iii) even deep optical surveys can provide 
relatively high spectral resolution, enabling detailed insights into the 
velocity profile of the absorbing gas, as well as its position relative to 
the Milky Way's disk, central region, and large-scale environment - kinematic 
information that other methods cannot provide; iv) when combined with 
observations from other probes, such as Fast Radio Bursts (FRBs), these data 
may help to constrain the iron abundance in the Milky Way's hot corona and shed 
light on the enrichment history of this gas.
\citet{zastrocky2018} stacked KECK/HIRES spectra from the 
KODIAQ sample to provide a first upper column-density limit for Fe\,{\sc xiv}
in the Milky Way halo; their limit is, however, more than two orders of magnitude 
above the expected level.

Motivated by the many upcoming spectroscopic galaxy surveys from other
instruments, which will provide millions of medium- to high-resolution 
spectra, we revisit this idea and provide for the first time
testable predictions for the expected stengths of the [Fe\,{\sc x}] 
$\lambda 6374.5$ and [Fe\,{\sc xiv}] $\lambda 5302.9$ absorption lines 
from the Galaxy's hot coronal gas based on models, simulations, and 
observations of the distribution of the CGM in the Milky Way halo.

This paper is organized as follows: in Sect.\,2, we review the physical properties
of the [Fe\,{\sc x}] $\lambda 6374.5$ and [Fe\,{\sc xiv}] $\lambda 5302.9$ lines
and discuss the results from previous attempts of detecting them. 
In Sect.\,3, we discuss the hot gas distribution in the Milky Way halo and constrain
the expected hot gas column density from an analytic model, from the HESTIA
simulations of the Local Group, and from dispersion measurements. 
In Sect.\,4, we make predictions for the expected 
absorption strength of these two lines and provide upper limits for the column
densities of log $N$(Fe\,{\sc x}) and log $N$(Fe\,{\sc xiv}) from a combined
stack of VLT/UVES and KECK/HIRES. We discuss the prospects of detecting
[Fe\,{\sc x}] $\lambda 6374.5$ and [Fe\,{\sc xiv}] $\lambda 5302.9$ absorption
in stacked spectra from upcoming spectroscopic surveys in Sect.\,5.
Summary and conclusions are given in Sect.\,6.


\begin{figure}[t!]
\begin{center}
\resizebox{0.85\hsize}{!}{\includegraphics{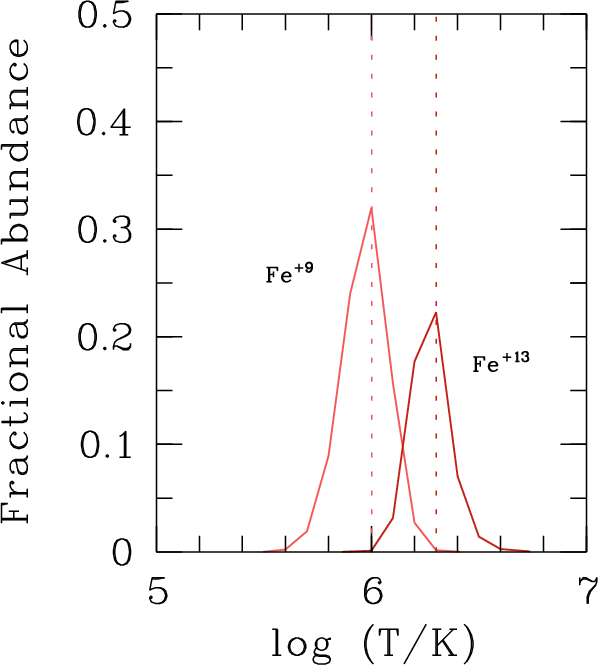}}
\caption[]{
Fractional abundances of Fe$^{+9}$ and Fe$^{+13}$ in
a collisional ionization equilibrium \citep[from][see also Sect.\,2.1]{bryans2009}.
}
\end{center}
\end{figure}


\section{Tracing million degree gas with [Fe\,{\sc x}] and [Fe\,{\sc xiv}] absorption}

\subsection{Constraints from atomic physics}

The optical intersystem transitions of [Fe\,{\sc x}] at $\lambda_0=6374.51$ \AA\,
and [Fe\,{\sc xiv}] at $\lambda_0 =5302.86$ \AA\, \citep{shirai2000}
\footnote{Throughout this paper, we provide air-wavelengths in units \AA\, for 
the optical transitions of [Fe\,{\sc x}] and [Fe\,{\sc xiv}], as given in the 
NIST Atomic Spectra Database \citep{kramida2023}, while we give vacuum wavelengths
for transitions of UV lines at $\lambda\leq 3000$ \AA.}
are magnetic dipole transitions (M1)
that are well known from observations of the solar corona, where these
transitions lead to prominent emission features: the ``coronal'' lines,
as known since 1939 \citep{grotrian1939}.
In contrast to the high density ($n_{\rm H}\sim 10^5$ 
cm$^{-3}$) coronal plasma of the Sun, the forbidden optical intersystem 
transitions of [Fe\,{\sc x}] and [Fe\,{\sc xiv}] can be detected in the hot,
low-density intergalactic and circumgalactic medium ($n_{\rm H}< 10^{-4}$
cm$^{-3}$, typically) only in absorption against extragalactic background
sources (such as quasars/quasi-stellar objects, QSOs, and other type of AGN).

The expected line strengths of [Fe\,{\sc x}] $\lambda 6374.5$ and
[Fe\,{\sc xiv}] $\lambda 5302.9$ absorption in a given hot-gas distribution 
depend on i) the total gas column density, ii) the local iron abundance
in the gas, iii) the fractional abundances of the 
Fe\,{\sc x} and Fe\,{\sc xiv} ions (e.g., dictated by the temperature in 
collisional ionization 
equilibrium), and iv) the transition probabilities (oscillator strengths)
of the $\lambda 6374.5$ and $\lambda 5302.9$ transitions. 
In the following, we discuss all of these parameters and their relation
to each other in preparation for our predictions of the [Fe\,{\sc x}] and
[Fe\,{\sc xiv}] line strengths in the hot coronal gas of the Milky Way
(Sect.\,4.1).

The total column density, $N_{\rm tot}\approx N$(H), of million degree 
coronal gas around the Milky Way (and other galaxies) has only partly 
been determined by X-ray observations (see introduction section), but 
can be estimated from analytic halo models and cosmological 
(magneto)hydrodynamical simulations as well as from measurements of the
dispersion measure against extragalactic Fast Radio Bursts.
This will be discussed in detail in Sect.\,3.

The local iron abundance in the gas, (Fe/H$)_{\rm Cor}$, can be expressed in terms 
of the solar iron abudance \citep[Fe/H$)_{\sun}=3.16\times 10^{-5}$;][]{asplund2009},
so that (Fe/H$)_{\rm Cor} \equiv X_{\rm Fe}$(Fe/H$)_{\sun}$ with $X_{\rm Fe}$
as free parameter to be defined.
The fractional abundances of Fe\,{\sc x} and Fe\,{\sc xiv} in the 
gas, $f_{\rm FeX}$ and $f_{\rm FeXIV}$, depend on the ionization state
of the medium (and is thus driven by the ionization mechanisms). 
It requires very high energies of $234$\,eV to ionize 
Fe$^{+8}$ to Fe$^{+9}$ and $361$\,eV from Fe$^{+13}$ to Fe$^{+14}$
\citep[e.g.,][]{sutherland1993}.

In the absence of an intense flux of extremely hard photons
(as expected in most astrophysical environments and,
in particular, in the CGM and IGM), collisional ionization at
million-degree temperatures is the only mechanism that is expected
to produce a significant population of Fe$^{+9}$ and Fe$^{+13}$
ions in diffuse intergalactic and circumgalactic gas. Assuming
collisional ionization equilibrium (CIE), and adopting
CIE models for Fe\,{\sc x} and Fe\,{\sc xiv} from \citet{bryans2009},
the maximum fractional abundance of Fe\,{\sc x} at its CIE peak
temperature ($T_{\rm p,FeX}=1.0\times 10^6$ K) is $p_{\rm FeX}=0.32$,
while for Fe\,{\sc xiv} it is $p_{\rm FeXIV}=0.22$ at
$T_{\rm p,FeXIV})=2.0\times 10^6$ K (see Fig.\,1).
Note that these fractional abundances do not change significantly
($\Delta p \leq 0.002$) if one considers non-equilibrium collisional 
ionization models, such as presented in \citet{gnat2007}.
With Einstein coefficients of $A_{\rm ki}=69.4$ s$^{-1}$ and 
$A_{\rm ki}=60.2$ s$^{-1}$ 
\citep{fuhr1988}, the transitions of
[Fe\,{\sc x}] $\lambda 6374.5$ and
[Fe\,{\sc xiv}] $\lambda 5302.9$ 
have extremely small oscillator strengths, $f_{\lambda 6374}=2.1 \times
10^{-7}$ and $f_{\lambda 5302}=5.1 \times 10^{-7}$. 

For a weak, unsaturated absorption line of an ion $Y$, there is a simple, 
linear relation between the observed equivalent width, $W_{\lambda}$, 
and the ion column density, $N(Y)$, in the gas:
 \begin{equation}
 \left(\frac{W_{\lambda}}{\rm m\textrm{\AA}}\right)=
 8.85\times 10^{-18}\,f\,\left(\frac{N(Y)}{{\rm cm}^{-2}}\right)
 \left(\frac{\lambda_0}{\rm \textrm{\AA}}\right)^2.
 \end{equation}
Here, $\lambda_0$ denotes the laboratory wavelength of the transition and
$f$ its oscillator strength. 

At their respective CIE peak temperatures, the column densities 
of Fe\,{\sc x} and Fe\,{\sc xiv} then can be expressed as 
$N$(Fe\,{\sc x}$)=p_{\rm FeX}\,X_{\rm Fe}\rm{(Fe/H)_{\sun}}$\,$N({\rm H})=
1.01\times 10^{-5}\,X_{\rm Fe}$\,$N({\rm H})$ and
$N$(Fe\,{\sc xiv}$)=p_{\rm FeXIV}\,X_{\rm Fe}\rm{(Fe/H)_{\sun}}$\,$N({\rm H})=
6.95\times 10^{-6}\,X_{\rm Fe}$\,$N({\rm H})$, so that
 \begin{equation}
 \left(\frac{W_{6347}}{\rm m\textrm{\AA}}\right)=0.0763\,X_{\rm Fe}\,
 \left(\frac{N({\rm H})}{10^{20}\,{\rm cm}^{-2}}\right)    
 \end{equation}
and
 \begin{equation}
 \left(\frac{W_{5302}}{\rm m\textrm{\AA}}\right)=0.0882\,X_{\rm Fe}\,
 \left(\frac{N({\rm H})}{10^{20}\,{\rm cm}^{-2}}\right).
 \end{equation}

We see that for a 1 m\AA\, absorption line to arise in the
[Fe\,{\sc x}] $\lambda 6374.5$ ([Fe\,{\sc xiv}] $\lambda 5302.9$) line
it requires a total ionized gas column density as large as 
$1.3\times 10^{21}$ cm$^{-2}$ ($1.1 \times \times 10^{21}$ cm$^{-2}$)
at a solar Fe iron abundance. 

Because of the high gas temperature, the resulting absorption 
lines will be substantially broadened.
The expected thermal Doppler parameter is
 \begin{equation}
 b_{\rm th}=\sqrt{\frac{2kT}{m_{\rm ion}}} 
 \approx 0.129\times\sqrt{\frac{T/{\rm K}}{A_{\rm ion}}}
 = 17\,{\rm km\,s}^{-1}\,\left(\frac{T}{10^6\,{\rm K}}\right).
 \end{equation}

In this equation, $k$ is the Boltzmann constant, $m_{\rm ion}$ is 
the ion mass, and $A$ is the relative atomic mass of the absorbing ion. 
In case of Fe, $A_{\rm Fe}=55.815$ \citep[see NIST Atomic Spectra 
Lines Database;][]{kramida2023},
so that for a gas temperature of $T=10^6$ K ($2\times 10^6$ K) we 
have $b_{\rm th}=17$ km\,s$^{-1}$ ($24$ km\,s$^{-1}$).
Note that because of the relatively high mass of iron, this $b_{\rm th}$ value is
substantially smaller than the $b_{\rm th}$ values expected for 
thermally broadenend H\,{\sc i} Ly\,$\alpha$ absorbers in the CGM
\citep[$A_{\rm H}=1$;][]{richter2020}. The expected small 
breadth of the Fe lines favours their detectability in a typical AGN 
continuum spectrum that is altered with noise features (see Fig.\,2).

Because $b_{\rm th}$ is predicted to be relatively small for the
Fe lines, non-thermal broadening mechanisms, e.g., from the differential 
co-rotation of the coronal gas with the disk and bulk 
motions of the hot gas, are expected to be important. Such non-thermal 
motions are commonly parameterized with $b_{\rm nth}$, so that the 
expected observed Doppler parameter $b$ is composed of both of these 
components, $b^2=b_{\rm th}^2+b_{\rm nth}^2$.
The actual contribution of $b_{\rm nth}$ to $b$ is challenging to
estimate without knowing the systematic large-scale motions and the 
internal velocity dispersion of the hot gas in the corona.

There is observational evidence that the hot corona is partly 
co-rotating with the underlying disk \citep{hodges2016}. The implication 
of such a systematic motion along a line of sight from the Sun to 
Milky Way's virial radius for the shape of the Fe absorption lines 
is unclear, however, as it remains to be determined how the rotation speed scales
with the vertical height of the hot gas layers.
The role of non-thermal gas motions for $b$ will be 
be further discussed in Sect.\,3.2, where we analyze the velocity
dispersion of the coronal gas in the HESTIA simulations.

For a given equivalent width, the central absorption depth, $\cal{D}$, of 
an absorption line depends on its width and thus on $b$. Following 
\citet{richter2020}, we can write (assuming a Gaussian shape):
 \begin{equation}
 {\cal D}=0.139\,\left[ \frac{W_{\lambda}} {\rm m\textrm{\AA}} \right]\,
 \left[\frac{b}{{\rm km\,s}^{-1}}\right]^{-1}.
 \end{equation}

\subsection{Previous [Fe\,{\sc x}] and [Fe\,{\sc xiv}] observations}

In the light of the values obtained in the previous sub-section, we now
discuss the results from previous studies of [Fe\,{\sc x}] and
[Fe\,{\sc xiv}] absorption in nearby hot gas and towards the Magellanic Clouds.

Already 40 years ago, \citep{hobbs1984a,hobbs1984b,hobbs1985,pettini1986}
searched for [Fe\,{\sc x}] and [Fe\,{\sc xiv}]
absorption in the local Milky Way ISM 
using high S/N spectra of nearby stars. These studies
did not find any convincing signal down to a level of $\sim 1$ m\AA.
However, million-degree gas within the Milky-Way gas disk is expected to reside
only in localized enviroments at sub-kpc scales (e.g., the local hot bubble 
or other supernova bubbles and chimneys) that do not provide sufficient
column density for [Fe\,{\sc x}] and [Fe\,{\sc xiv}] absorption to be detected
at a $\sim 1$ m\AA\, level \citep[e.g.,][]{liu2017}. This is because for a hot-gas 
density as high as $n_{\rm H}=10^{-2}$ cm$^{-3}$, it requires an absorption path length 
of 3.2 kpc to achieve a hydrogen column density of $N$(H$)=10^{20}$ cm$^{-2}$.
Therefore, the reported non-detections in the above-listed studies are consistent
with equations (2) \& (3).

From the analysis of the optical spectrum of the extremely bright supernova 
SN\,1987A in the LMC, several groups \citep{wang1989,pettini1989,malaney1988}
reported the detection of a weak ($W_{\lambda}\sim 3$ m\AA) absorption feature 
near $6380$ \AA\, that was identified as Doppler-shifted [Fe\,{\sc x}] $\lambda 6374.5$ 
absorption arising from hot gas in the direction of the LMC \citep[][]{pettini1989}. 
If true, this would imply  a large column of ionized gas in this direction,
log $N$(H\,{\sc ii})$\approx21.5$ (Eq.\,2), indicating the existence of a huge
reservoir of baryonic matter hidden in hot gas. The interpretation of the
observed absorption feature near $6380$ \AA\, as [Fe\,{\sc x}]
absorption was, however, disputed by \citet{wampler1991}, who identified the
observed feature as a Doppler-shifted diffuse interstellar band (DIB). 
There are several weak DIBs and telluric lines present in the optical regions in which
the [Fe\,{\sc x}] $\lambda 6374.5$ and [Fe\,{\sc xiv}] $\lambda 5302.9$ lines
are located \citep[see, e.g.,][]{jenniskens1994}. The latter interpretation 
appears more likely, because the existence of a uniform layer of million-degree gas
around the LMC at a column density of log $N$(H\,{\sc ii})$\approx21.5$ would
imply an enormously high baryon-to-dark-matter ratio that is not in line with current
galaxy formation models. Such a large ionized gas column density is also not
supported by UV and X-ray observations of hot gas in the general direction of the LMC
\citep{krishnarao2022,locatelli2024}.

More recently, \citet{richter2014} discussed high S/N VLT/UVES observations 
of high-redshift QSOs as possible probes for [Fe\,{\sc x}] and [Fe\,{\sc xiv}] 
absorption in the Milky Way's coronal gas (e.g., their Fig.\,4). They also
proposed stacking techniques for UVES high-resolution and SDSS low-resolution
archival spectra to achieve the required S/N levels. From the stacking of 95 
pre-selected Keck HIRES spectra, \citet{zastrocky2018} derived $3\sigma$ upper 
limits for [Fe\,{\sc xiv}] absorption in the Milky Way halo of $W_{5302}\leq 7$ m\AA\, 
and $N$(Fe\,{\sc xiv}$)=5.5\times 10^{16}$ cm$^{-2}$ achieving a S/N of $\sim 300$.


\section{Expected properties of the Milky Way's hot coronal gas}

To determine realistic expectation values for $W_{6347}$ and $W_{5302}$ from
[Fe\,{\sc x}] and [Fe\,{\sc xiv}] absorption in the Milky Way's hot coronal gas,
we need to estimate $N$(H) for the coronal gas at the CIE peak temperatures
of Fe$^{+9}$ to Fe$^{+13}$ (Fig.\,1) in the range log $T=6.0-6.3$
(see Eqs.\,2 and 3) for extragalactic sightlines from the vantage point 
of the Sun. For this, we use i) the semi-analytic model of the hot gas 
distribution around low-redshift galaxies from \citet{richter2020}, ii) results 
from the constrained magneto-hydrodynamical simulation of the Milky Way and 
Local Group as part of the HESTIA project \citep{libeskind2020,damle2022,runger2025},
and iii) constraints from measurements of the dispersion measure of gas
in around the Milky Way using Fast Radio Bursts \citep{cook2023}.

\subsection{Constraints from a semi-analytic model}

In \citet{richter2020}, we used a semi-analytic approach to model the 
spatial density and temperature distribution of hot coronal gas of galaxies 
with virial halo masses in the range log $(M/M_{\sun})=10.6-12.6$
to predict strength, spectral shape, and cross section of the CBLAs as a 
function of galaxy-halo mass and line-of-sight impact parameter from an
external vantage point. In this model, it is assumed that the hot coronal
gas is confined in a dark-matter halo that follows a 
Navarro–Frenk–White (NFW) density profile \citep{navarro1995,klypin2001}.
After the initial collapse, the gas is assumed to be shock-heated to the 
halo's virial temperature and then cools within a characteristic cooling 
radius to become multi-phase \citep[see][for a detailed description of 
the original equations]{maller2004}.

To model the spectral signatures of such gas along QSO sightlines, we then 
have applied the {\tt halopath} code, as originally described in 
\citet{richter2012}. The same approach and modeling codes is used here to 
calculate $N$(H) for $T\geq10^{6}$ K in a Milky-Way mass halo from any given 
vantage point inside the halo by integrating $n_{\rm H}$ from the initial 
position out to the virial radius, $R_{\rm vir}$. 


\begin{figure*}[t!]
\begin{center}
\resizebox{0.9\hsize}{!}{\includegraphics{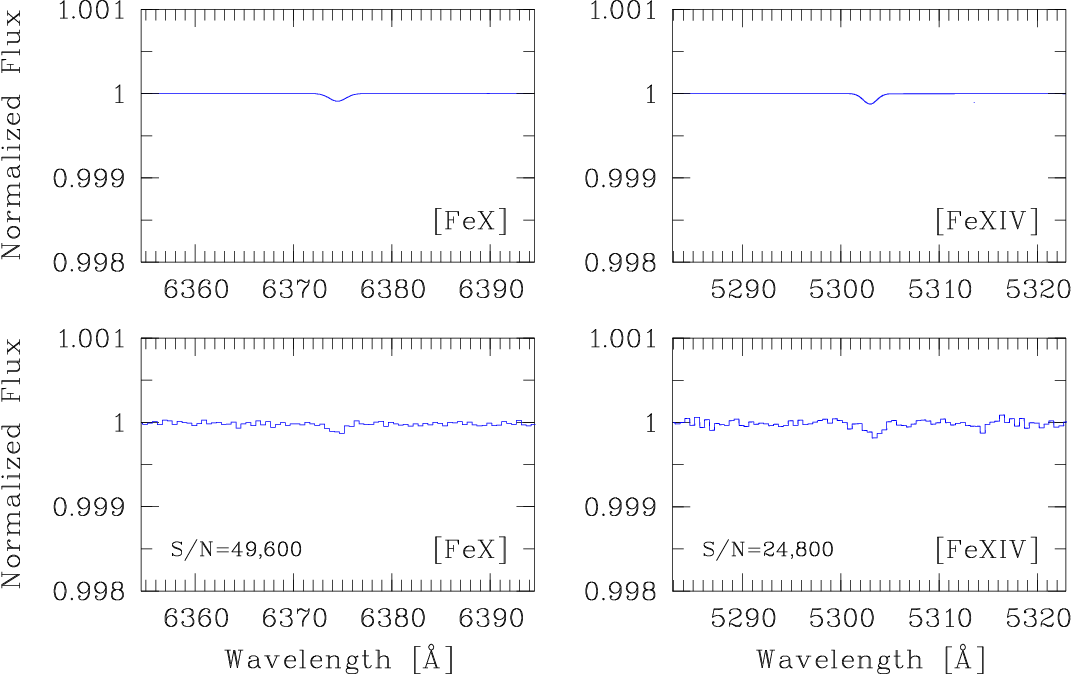}}
\caption[]{
Synthetic spectra of the [Fe\,{\sc x}] at $\lambda 6374.5$ and
[Fe\,{\sc xiv}] $\lambda 5302.9$ lines from our halo model
described in Sect.\,4.1. The upper two panels show the noise-free
absorption profiles, the lower two panels show the profiles with
the minimum S/N required for a $3\sigma$ detection, as given in the
lower left corners. The spectra are generated using a spectral
resolution of $R=45,000$ and a sampling of 1.5 pixels per resolution 
element.
}
\end{center}
\end{figure*}


For the Milky Way mode halo, we adopt a halo mass of log $(M/M_{\sun})=12.0$ 
and a virial radius of $R_{\rm vir}=250$ kpc and for sake of 
simplicity we assume the observer to be located in the geometrical
center of the spherical model halo
\citep[for more details about the model see][]{richter2020}.
If we only consider
gas with $T\geq10^6$ K, the total hydrogen column density 
of that gas between $R=0$ kpc and $R_{\rm vir}$ is typically
$N($H\,{\sc i}$)=1.0\times 10^{20}$ cm$^{-2}$. To put this number into context: a similar
column density is observed for neutral hydrogen in the
Magellanic Stream \citep[e.g.,][]{putman2003,fox2013,richter2013}, where the absorption
pathlength through the stream is substantially smaller (a few kpc, at most), but
the gas density is significantly higher (as is the neutral gas fraction) as well as
in the lowest $N$(H\,{\sc i}) directions in the Milky Way disk, such in the 
Lockman hole \citep{richter2001}.

\subsection{Constraints from the HESTIA simulations}

While the semi-analytic halo model discussed above provides important 
insights into the overall hot gas distribution in the CGM as a function 
of halo mass, it cannot deliver information on the kinematics of the 
absorbing gas cells, nor can it account for density and temperature fluctuations
in the hot CGM that are expected to arise due to the specific
cosmological environment of the host galaxy and because of feedback processes 
from supernovae and AGN activity.
Therefore, for a more realistic estimate of $N$(H) in the hot CGM
of a Milky-Way type galaxy, we consider magneto-hydrodynamical (MHD) high-resolution 
simulations of galaxies in a cosmological context.

We make use of the suite of simulations from the HESTIA (High-resolution 
Environmental Simulations of The Immediate Environment) project \citep{libeskind2020}
to systematically investigate the properties of the (simulated) Milky Way CGM
\citep{damle2022,biaus2022,runger2025}. HESTIA provides representative 
magneto-hydrodyanmical simulations of Milky Way-mass galaxies in their proper cosmographic 
environments using the AREPO moving-mesh code \citep{springel2010} and the AURIGA galaxy 
evolution model \citep{grand2017}. For details on HESTIA, see \citet{libeskind2020}.

We here use the three HESTIA simulations to estimate $N$(H) in the 
coronal gas from a Sun-like position at 8 kpc galactocentric distance within the (simulated) 
MW disk and pinpoint the velocity dispersion of the hot-gas pockets in the CGM 
along the various sightlines to constrain the non-thermal contribution 
to the Doppler parameter (Sect.\,2.1).
For each of the three MW simulations, we have constructed 100 randomly 
distributed sightlines from the position of the Sun that extend
to $R_{\rm vir}$ and derived $N$(H) by integrating over the
density distribution along each sightline. As for the semi-analytic model, 
we restrict our analysis to gas with temperatures $T\geq10^6$ K. The median 
value for $N$(H) from all 300 sightlines is $N$(H$)=5\times 10^{20}$ 
cm$^{-2}$.
Interestingly, this value is $\sim 5$ times higher than $N$(H) 
estimated from the simple semi-analytic model (Sect.\,3.1). The larger
column density of hot, ionized gas in the HESTIA simulation comes from
the important contributions of hot gas cells related to the Milky Way's (simulated) 
disk-halo interface, where supernova feedback adds to the energy and hot-baryon
budget of gas in the lower halo, and possibly from gas related to Milky Way satellites
within $R_{\rm vir}$ \citep[see also][]{damle2025}.

The (radial) velocity dispersions of the hot gas cells along the 300 sightlines 
vary between $30$ and $128$ km\,s$^{-1}$ with a median value of 
$\langle \sigma_v \rangle = 50$ km\,s$^{-1}$. We regard this latter value as a
characteristic velocity dispersion for the non-thermal gas motions along a 
typical halo sightline, which then dominates expected line width of the
[Fe\,{\sc x}] and [Fe\,{\sc xiv}] absorption (Sect.\,2.1).

\subsection{Constraints from Fast Radio Bursts} 

Another, independent constraint on the mass and extent of the hot gas in the
the Milky Way halo comes from FRBs \citep{lorimer2007}
and distant pulsars, using the well-known dispersive effect of charged particles 
on radio waves. By precisely measuring the arrival times 
of radio pulses (e.g., from pulsars or FRBs), the dispersive delay can 
be quantified using a quantity called the dispersion measure (DM).
Since free electrons by far dominate the dispersion
in a given ionized gas volume, the DM is proportional to the electron column 
density in the gas along a line of sight of length $L$, 
 \begin{equation}
 {\rm DM} = \int_0^L n_e dl = N(e^-).
 \end{equation}
And because hydrogen is the main electron donor in a fully ionized interstellar/intergalactic
plasma at solar or sub-solar metal abundance, we have $N(e^-)\approx N$(H), so that 
the DM is a direct measure of the total (ionized) gas mass along a line of sight.

DM measurements deliver information on the total ionized gas column 
towards a background FRB or pulsar. They do not, however, provide any information
on how the particles are distributed along the line of sight. To localize
hot gas in and around the Milky Way and estimate its mass, the combination 
of many FRB and/or pulsar sightlines of different lengths is required
\citep[see, e.g.,][]{prochaska2019}.
From a sample of 93 FRBs for Galactic latitudes $|b|\geq 30$ deg from the 
CHIME/FRB project, \citet{cook2023} derived a range of DM$=88-141$ pc\,cm$^{-3}$.
These values correspond to a range in total ionized gas column densities of
$N$(H)$=2.7-4.4 \times 10^{20}$cm$^{-2}$, thus slightly below the value
estimated from the HESTIA simulations, but clearly above the one 
derived from the semi-analytic model.


\section{Observational limits}

\subsection{Expected line strengths and profiles}

If we assume a total hydrogen column density in the Milky Way's hot coronal
gas of $N$(H$)=5\times 10^{20}$ cm$^{-2}$ and an iron abundance in the corona of 
half of the solar value \citep{martynenko2022}, the expected line strengths from 
equations (2) and (3) come out as 
$W_{6374}=190\,\mu$\AA\,and $W_{5302}=220\,\mu$\AA.
The corresponding Fe\,{\sc x} and Fe\,{\sc xiv} column densities are
log $N$(Fe\,{\sc x}$)=15.40$ and log $N$(Fe\,{\sc xiv}$)=15.23$.
For a hot-gas temperature of $2\times 10^6$ K and non-thermal gas motions 
characterized by $b_{\rm nt}=50$ km\,s$^{-1}$ (from the HESTIA simulations, see above),
the expected total Doppler parameter is $b=58$ km\,s$^{-1}$.

To visualize the optical appearance of such weak spectral features, we have
generated synthetic spectra of the [Fe\,{\sc x}] $\lambda 6374.5$ and
[Fe\,{\sc xiv}] $\lambda 5302.9$ lines in Fig.\,2 using a spectral
resolution of $R=45,000$ and a sampling of 1.5 pixels per resolution element. The upper
two panels show the blank (noise-free) absorption profiles, for the
lower two panels we have added noise at a level, that a $3\sigma$ detection
of these features is realized. As we can see, a S/N of $\sim 50,000$ is required
to significantly detect [Fe\,{\sc x}] absorption at the expected level,
for [Fe\,{\sc xiv}] it is $\sim 25,000$. 


\begin{figure}[t!]
\begin{center}
\resizebox{0.85\hsize}{!}{\includegraphics{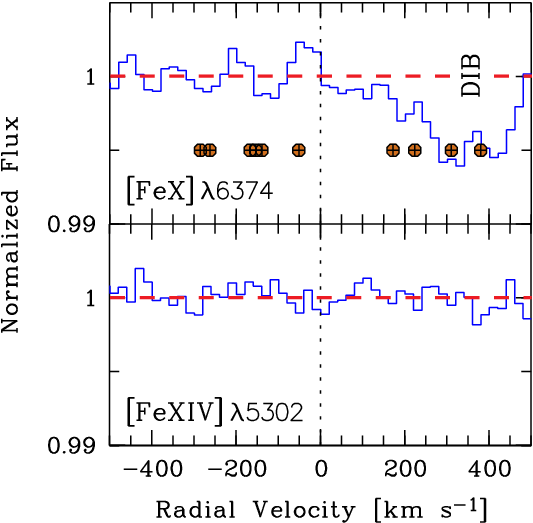}}
\caption[]{
Velocity profiles from stacks of the [Fe\,{\sc x}] $\lambda 6374.5$ and
[Fe\,{\sc xiv}] $\lambda 5302.9$ lines from a combined sample of 186/156 VLT/UVES
and KECK/HIRES spectra (see Sect.\,4.2). The resulting S/N per $20$ km\,s$^{-1}$
pixel element is $900$ for the stacked [Fe\,{\sc x}] profile and $1,240$ for
[Fe\,{\sc xiv}]. In the profile of the [Fe\,{\sc x}] line, a weak DIB is detected
near $+300$ km\,s$^{-1}$. The positions of telluric features in the [Fe\,{\sc x}]
profile are indicated with the crossed circles \citep[from][]{pettini1989}.
Note that the S/N ratios achieved here still are $1-2$ orders of magnitude 
below the values predicted by our models (see Fig.\,2).
}
\end{center}
\end{figure}


\subsection{Stacking of archival high-resolution optical AGN spectra}

Even with the most efficient spectrographs on $8-10$m class telescopes, 
it is unrealistic to achieve the required S/N of $>20,000$ for a medium- to 
high-resolution ($R\ge 15,000$) optical spectrum of an individual extragalactic 
background source within reasonable integration times. 
For the UVES spectrograph, for instance, \citet{dodorico2016} and 
\citet{kotus2017} presented high-resolution ($R\geq 45,000$) spectra of 
the high-redshift QSOs HE\,0940$-$1050 and HE\,0515$-$4414, respectively. 
These data have S/N ratios of $\sim 500-600$ per resolution 
element and are based on many individual 
pointings totaling several dozen hours of integration time with the VLT.
To our knowledge, there are no other VLT/UVES (or KECK/HIRES) spectra of 
an individual QSO that provide a S/N significantly higher than these values.

Because every extragalactic sightline passes the Milky Way halo, however,
stacking clearly is the most promising approach to achive S/N ratios larger 
than 1000 for a Milky Way composite spectrum. Using low-resolution 
($R\sim 2,200$) SDSS spectral data from stars, galaxies and quasars, 
\citet{lan2015} and \citet{murga2013} have 
achieved a S/N of several thousand by stacking a few hundred thousand
SDSS spectra to investigate interstellar DIBs and Ca\,{\sc ii} absorption 
in the Milky Way foreground gas. They did not, however, find evidence for 
[Fe\,{\sc x}] and [Fe\,{\sc xiv}] absorption in their stacked data set
\citep[Menard, priv.\,comm.; see also][their Fig.\,2]{lan2015}, which is 
in line with the predicted strength of the absorption as calculated above.
We also note that the spectral resolution of SDSS ($R=2200-2850$, equivalent
to $\Delta v\geq 136$ km\,s$^{-1}$) is not sufficient 
to resolve the two [Fe\,{\sc x}] and [Fe\,{\sc xiv}] lines at the 
expected widths.

\citet{zastrocky2018} used a sample of 95 KECK/HIRES spectra from
the KODIAQ survey \citep{omeara2017} for a spectral stack to search
for [Fe\,{\sc xiv}] absorption in the Milky Way halo. They achieve a 
S/N of $\sim 300$ and a column density limit of log $N$(Fe\,{\sc xiv}$)\leq 16.78$
from the mean stack of their composite spectrum, which is far below
(two orders of magnitude in terms of S/N) the required sensitivity.

For this paper, we have combined the KECK/HIRES spectra from the KODIAQ 
data basis with VLT/UVES spectra from the SQUAD project \citep{murphy2019} and other archival
VLT/UVES data. Our goal is to boost the detection limit for [Fe\,{\sc x}] and [Fe\,{\sc xiv}] absorption 
from a high-resolution composite spectrum that is based on high-quality data from two of 
the most powerful optical spectrographs installed on $8-10$m class telescopes currently
available. Both spectral data sets have similar spectral resolution
$R\sim 30,000-60,000$, but the average S/N in the VLT/UVES spectral data is 
$2-3$ times higher than for the KECK/HIRES data including some spectra with
exceptionally high S/N (see above) from the monitoring of AGN variability.
Note that we have previously used VLT/UVES spectra from the SQUAD project
to systematically investigate Ca H \& K absorption in the Milky Way halo 
\citep{richter2005,benbekhti2008,benbekhti2012}, in the low-redshift IGM \citep{richter2011},
and in selected Damped Lyman-$\alpha$ absorbers \citep{guber2016,guber2018}.
We therefore are highly familiar with the data quality issues of the 
SQUAD database.

For the stacking, we developed a custom stacking routine that (in a 
fully automated fashion) identifies blending features from intervening absorbers in 
the spectral regions near 5300 and 6375 \AA, performs data quality checks, and 
determines the local S/N. Using pre-defined selection criteria and bin sizes, the code
then bins the data and stacks the selected spectra, where each spectrum is weighted by the 
inverse variance from the noise statistics in the regions around the two Fe lines. 
In our case, the initial sample contains 439 UVES and 300 HIRES spectra (i.e., 739 in total). 
Before stacking, we pre-selected spectra that i) have no contaminating 
blending line within $\pm 500$ km\,s$^{-1}$ of the laboratory wavelength of the 
[Fe\,{\sc x}] and [Fe\,{\sc xiv}] lines, and ii) that (each) have a S/N$\geq 20$ per
resolution element. This selection leaves us with 183 spectra for [Fe\,{\sc x}] and 156 
spectra for [Fe\,{\sc xiv}] to carry out the stacking procedure.
Before stacking, the indivdual spectra were re-binned to achieve a homogeneous 
sampling, where the bin size was chosen to be $20$ km\,s$^{-1}$, sufficient to resolve an absorption
line that has an expected Doppler parameter of $b\geq 50$ km\,s$^{-1}$. Finally, the stacked 
spectrum was normalized in the regions of interest using a linear continuum fit.
The resulting composite velocity profiles for the [Fe\,{\sc x}] and [Fe\,{\sc xiv}] lines
after stacking are shown in Fig.\,3. 

For [Fe\,{\sc xiv}] (lower panel), the stacked composite at $5302$ \AA\, is featureless 
and achieves a very high S/N$=1,240$ per $20$ km\,s$^{-1}$ wide pixel element, which 
translates into a $3\sigma$ upper equivalent-width limit of $W_{5302}\leq 0.9$ m\AA\, 
and log $N$(Fe\,{\sc xiv}$)\leq 15.85$. 

For [Fe\,{\sc x}] (upper panel), the composite spectrum is slightly noisier 
between $-500$ and $+200$ km\,s$^{-1}$ (S/N$=900$), corresponding to a $3\sigma$ upper 
equivalent-width limit of $W_{6374}\leq 1.4$ m\AA\, and log $N$(Fe\,{\sc x}$)\leq 16.27$.
There are a number of telluric features known in the spectral region between $6365$ and
$6385$ \AA\, \citep[see Table 2 in][]{pettini1989} that could potentially contaminate
any [Fe\,{\sc x}] $\lambda 6374.5$ absorption. Their positions are indicated in Fig.\,3.
The [Fe\,{\sc x}] profile also shows an absorption feature at a $\sim 0.5$ percent absorption-depth level 
between $+250$ and $+450$ km\,s$^{-1}$. This feature represents a DIB at $6379.32$ \AA\, 
\citep{jenniskens1994} and is well-known from Galactic stellar sightlines with high extinction 
\citep[e.g.,][their Fig.\,3]{hobbs2008}. It is also identified at a $\sim 0.5$ percent level 
in a stack of more than 40,000 SDSS spectra of Galactic stars along sightlines with 
$E(B-V)>0.1$ mag \citep{lan2015}.

In our [Fe\,{\sc x}] composite spectrum, we measure an equivalent width of $29.5$ m\AA\,
for this feature. The fact that this DIB emerges in our stacked composite spectrum of 
extragalactic AGN sightlines, which are predominantly located at high Galactic latitudes,
strongly supports the conclusion that the feature near $6380$ \AA\, in the spectrum of SN1987A 
in the LMC is caused by the DIB at $6379.32$ \AA\, and not by [Fe\,{\sc x}] absorption 
from hot coronal gas (see discussion in Sect.\,2.2). The detection of this feature in our
composite spectrum also demonstrates the robustness of our stacking procedure.

In conclusion, our stacking experiment suggests that the [Fe\,{\sc x}] 
$\lambda 6374.5$ line in composite spectra is usable only within a limited velocity 
range of $-50$ and $+150$ km\,s$^{-1}$ due to potential blending with a DIB and 
telluric features. In contrast, the [Fe\,{\sc xiv}] $\lambda 5302.9$ line offers 
a more promising alternative, as it is unaffected by such blending, allowing the 
full velocity range of the Milky Way to be explored.


\section{Outlook}

\subsection{General prospects from future spectroscopic surveys}

While the current archival spectral data base is insufficient to reach the required
S/N to obtain meaningful constraints on [Fe\,{\sc x}] and [Fe\,{\sc xiv}] in the Milky Way's
coronal gas via stacking, the prospects from observational surveys in the near future are 
excellent. With the advent of several new survey instruments and the installation of extremely 
sensitve optical telescopes such as the ESO Extremely Large Telescope (ELT; {\tt eso.elt.org}), 
there soon will be several millions of 
spectra available at medium- to high spectral resolution that will open completely new
windows to search for extremely weak absorption features of various ions in the foreground
gas in the Milky Way's interstellar and circumgalactic medium \citep{dodorico2024}.

\citet{fresco2020} have explored in detail the prospects of using future spectroscopic 
surveys to investigate hot gas at $\sim 10^7$ K around damped Ly\,$\alpha$ absorbers
in the early Universe by stacking a large number of optical spectra. Their goal is to 
put constraints on the absorption strength of the red-shifted forbidden [Fe\,{\sc xxi}] 
$\lambda 1354.1$ UV line \citep{anderson2016} in the hot gas around these 
(proto-)galactic structures at 
median redshift of $\langle z_{\rm abs} \rangle =2.5$. In their paper, the authors
discuss the instrumental capabilities of future spectroscopic surveys with millions of 
optical spectra from projects including 
i) the Multi-Object Spectrograph Telescope (4MOST), a high-multiplex spectroscopic
survey instrument for the 4m VISTA telescope at ESO \citep{dejong2019,mainieri2023},
for which science operations expected to starting in mid-2026;
ii) the WEAVE instrument \citep{jin2024}, designed for the William Herschel Telescope (WHT), which 
had first light in Dec 2022, and 
iii) the AmazoNes high Dispersion Echelle Spectrograph (ANDES; formerly named HIRES), 
which will be installed on 39m Extremely Large Telescope (ELT) currently under
construction \citep{marconi2022,dodorico2024}.
For details of these instruments and their performance in the context of stacking
experiments we refer to the discussion in \citet{fresco2020}.

The search for redshifted [Fe\,{\sc xxi}] absorption around high-$z$ DLAs, such
as described in \citet{fresco2020}, is not only limited by the number of high-redshift 
($z>2.2$) background quasars, but also by the cosmological number density (e.g., cross-section) 
of these absorbers. By contrast, every extragalactic sightline passes the hot coronal gas of the 
Milky Way, boosting the number of available spectra for stacking experiments using 
[Fe\,{\sc x}] and [Fe\,{\sc xiv}] and other weak absorption features in the Galaxy's 
CGM and ISM. 


\begin{table}[t!]
\begin{center}
\caption[]{Predictions for detecting [Fe\,{\sc xiv}] in future surveys}
\begin{normalsize}
\begin{tabular}{rr}
\hline
$\langle$ S/N $\rangle ^{\rm a}$ & ${\cal N}_{\rm stack}$$^{\rm b}$ \\
\hline
 10 & $6.3\times 10^6$ \\
 25 & $1.0\times 10^6$ \\
 50 & $2.5\times 10^5$ \\
100 & $6.3\times 10^4$ \\
\hline
\end{tabular}
\noindent
\\
$^{\rm a}$\,Average S/N per $50$ km\,s$^{-1}$ wide \\
bin element in individual spectra; \\
$^{\rm b}$\,Number of usable spectra for stacking.
\end{normalsize}
\end{center}
\end{table}


Achieving a S/N as high as $25,000$ in a spectral stack, as estimated for the [Fe\,{\sc xiv}]
line in Sect.\,4.2, will be extremely challenging, however. Depending on the type
of background sources (e.g., AGN, galaxy, star) and their redshift distribution,
only a certain fraction of the available spectra can be used for a stack in the
[Fe\,{\sc x}]  and [Fe\,{\sc xiv}] wavelength region due to blending effects and 
continuum issues. For our VLT/UVES sample described in Sect.\,4.2, for example, this 
fraction comes out to $20-25$ percent. 

In Table 1, we list the minimum number of usable spectra, ${\cal N}_{\rm stack}$,
required to reach a S/N of $25,000$ per $50$ km\,s$^{-1}$ wide bin element (second column) 
as a function of the average S/N per bin in the spectral sample. 
For a reliable estimate for ${\cal N}_{\rm stack}$, one needs to know the overall sample size
of available sources, their exact type and redshift distribution as well as 
their S/N distribution. The actual S/N achieved from a mix of possible instruments
and background sources would also need to be evaluated from a large set
of mock spectra with properties similar to those of the real data.

\subsection{Example: potential stacking experiments with 4MOST data}

In the following, we elaborate more on the scientific potential of up-coming spectral surveys
for [Fe\,{\sc xiv}] stacking experiments by focussing on the 
4MOST instrument \citep{dejong2019}. Designed as a high-multiplex spectroscopic
survey instrument for the 4m VISTA telescope at ESO, 4MOST has both a 
Low-Resolution Spectrograph (LRS; $R\approx 5,600$ at $5302$ \AA, corresponding 
to $\Delta v=53.5$ km\,s$^{-1}$) and a High-Resolution Spectrograph (HRS;
$R\approx 19,000$ at $5302$ \AA, corresponding to $\Delta v=15.8$ km\,s$^{-1}$).
In principle, the spectral data from both of these instrumental setups therefore can
be used for [Fe\,{\sc xiv}] stacking experiments.
4MOST is expected to deliver LRS and HRS data for several million extragalactic
sources (AGN and galaxies) as part of the Extragalactic Consortium Surveys 
\citep[e.g.][]{richards2019,merloni2019} and the Extragalactic Community Surveys 
\citep[e.g.][]{bauer2023,peroux2023} covering a broad range of S/N ratios.
With such a rich data set, it hopefully will
be possible to constrain the mean [Fe\,{\sc xiv}] column density in the Milky Way
disk and halo and estimate the total amount of million-degree gas in our Galaxy.

As mentioned above, the biggest advantage of using stacked optical spectra compared 
to X-ray observations and studies of the dispersion measure
lies in the (relatively) high spectral and radial-velocity resolution of optical 
spectrographs (even in their ``low-resolution'' modes). Also a moderate
velocity resolution of $\Delta v \approx 50$ km\,s$^{-1}$ will be sufficient
to explore the radial velocity profile of the hot gas in and around
the Milky Way and to estimate its mass as a function of distance to the Sun.

If sufficient S/N can be achieved even with spectral sub-samples, it may also
be possible to investigate the patchiness of the hot gas in the Milky Way
on angular scales \citep[see][]{kaaret2020} and - together with the velocity
information- to localize the hot gas reservoirs in the inner and outer Milky Way halo, such in
the disk-halo interface \citep{savage2003} and in the extra-planar regions influenced by the 
Galaxy's nuclear ouflow \citep[``eROSITA bubbles'',][]{predehl2020,sarkar2024}. 
Such data may also provide
important information on the possible co-rotation of the hot halo with the underlying 
disk, as proposed by \citet{hodges2016} based on low-resolution X-ray data.

Because of its relatively
large angular extent \citep[][their Fig.\,8]{richter2017a} and
its substantial radial velocity offset from the Milky Way ($300$ km\,s$^{-1}$),
also the extended corona of M31 might be a potential target for a [Fe\,{\sc xiv}] 
stacking experiment using extragalactic sightlines in the direction of the
Local Group barycenter \citep[see also][]{lehner2015,lehner2020}. Such
observations also could be coupled to FRB measurements to constrain
$N$(Fe)/$N$(H) and thus the metallicity of the coronal gas in the 
Milky Way and M31 haloes.

While 4MOST will also survey hundred throusands of LMC and SMC stars
and Milky Way halo stars \citep{cioni2019,helmi2019,christlieb2019}, it
is impossible to forecast their suitability for [Fe\,{\sc xiv}]
stacking experiments to constrain the hot-gas mass within the inner 
$\sim 50$ kpc of the Milky Way halo and the hot halo of the Magellanic 
Clouds \citep{lucchini2024} because of the complex stellar continua.


\section{Summary and conclusions}

In this paper, we have explored the possibility of studying the hot coronal
gas of the Milky Way at $T\geq 10^6$ K by analyzing the highly forbidden solar coronal 
lines of [Fe\,{\sc x}] and [Fe\,{\sc xiv}] in the optical in absorption against
bright extragalactic background sources. The main results of our study can be summarized 
as follows:\\
\\
(1) Using an analytic model of the Milky Way's coronal gas distribution,
results from the HESTIA simulations of the Local Group, and constraints from 
FRB observations, we predict that the
mean hydrogen column density for $T\geq 10^6$ K gas along a 
typical halo sightline from the Sun to 
the Galaxy's virial radius is $N$(H)$\approx 5\times 10^{20}$ cm$^{-2}$. Assuming the
gas to be in a collisional ionization equilibrium and adopting an iron abundance
half of the solar value, the expected Fe\,{\sc x} and Fe\,{\sc xiv}
column densities are log $N$(Fe\,{\sc x}$)=15.40$ and
log $N$(Fe\,{\sc xiv}$)=15.23$, where we expect non-thermal gas motions 
to dominate the Fe\,{\sc x} and Fe\,{\sc xiv} line widths.\\
\\
(2) Based on the atomic data for the lines of [Fe\,{\sc x}] $\lambda 6374.5$ and
[Fe\,{\sc xiv}] $\lambda 5302.9$, we predict characteristic equivalent
widths for the coronal gas of $W_{6347}=190 \mu$\AA\,and $W_{5302}=220 \mu$\AA.
We investigate the typical velocity dispersion of the hot gas cells in the HESTIA
simulation and from this generate synthetic absorption profiles of these extremely 
weak lines with and without noise. We conclude that a minimum S/N of $\sim 50,000$
is required to detect [Fe\,{\sc x}] $\lambda 6374.5$ absorption at a $3\sigma$ 
level, while for [Fe\,{\sc xiv}] $\lambda 5302.9$ the required S/N is $\sim 25,000$. \\
\\
(3) We combine archival high-resolution spectral data of 739 AGN from VLT/UVES and 
KECK/HIRES to construct a composite spectrum from stacking and to provide
upper limits for $N$(Fe\,{\sc x}) and $N$(Fe\,{\sc xiv}) in the Milky Way halo.
As expected, we do not find [Fe\,{\sc x}] and [Fe\,{\sc xiv}] absorption in the 
stacked spectrum, but obtain $3\sigma$ upper limits of log $N$(Fe\,{\sc x}$)\leq 16.27$ and
log $N$(Fe\,{\sc xiv}$)\leq 15.85$. Our composite spectrum achieves a S/N 
$\approx 1,240$ near $5302$ \AA\, and $\approx 900$ near $6374$ \AA. A feature near 
$6380$ \AA\,is clearly evident in the composite spectrum, the measured equivalent
width being $29.5$ m\AA. We identify this feature as a DIB in the foreground
interstellar gas in the Milky Way disk.\\
\\
(4) Finally, we discuss the possibilty of detecting [Fe\,{\sc x}] and [Fe\,{\sc xiv}]
absorption using new instruments and on-going/future observational campaigns.
Our study implies that the up-coming extragalactic spectral surveys with millions 
of medium- to high-resolution optical spectra
will provide an unprecendent wealth of absorption-line data that holds the prospect
of detecting the extremely weak [Fe\,{\sc x}] and [Fe\,{\sc xiv}] lines in the Milky Way's
coronal gas, providing important information on the hot component of the Galaxy's CGM
and the total mass therein. With a somewhat higher oscillator strength, a spectral
region not being contaminated by DIBs and telluric features, and a peak abundance near 
$2\times 10^6$ K, the [Fe\,{\sc xiv}] $\lambda 5302.9$ line represents the most
promising candiate to explore the coronal gas of the Milky Way with the ``coronal
lines'' of iron.


\begin{acknowledgements}

This study makes extensive use of the SQUAD and KODIAQ spectral data libraries;
we thank M.T. Murphy and J.M. O'Meara for sharing their data with us and providing
helpful comments.

\end{acknowledgements}


\end{document}